\documentclass{webofc}
\usepackage[varg]{txfonts}   
\begin{document}
\title{Charm-baryon enhancement and charm fragmentation fractions in small systems measured with ALICE}
%
%

\author{\firstname{Jianhui} \lastname{Zhu}\inst{1,2}\fnsep\thanks{\email{jianhui.zhu@cern.ch}} for the ALICE Collaboration
}

\institute{Institute of Particle Physics, Central China Normal University, 152 Luoyu Road, 430079 Wuhan, China
\and
           GSI Helmholtz Centre for Heavy Ion Research, Planckstraße 1, 64291 Darmstadt, Germany
          }

\abstract{%
Recent measurements of charm-baryon production at midrapidity by the ALICE collaboration show baryon-to-meson yield ratios significantly higher than those measured in $\rm e^+e^-$ collisions, suggesting that the charm fragmentations are not universal across different collisions systems. Thus, measurements of charm-baryon production are crucial to study the charm quark hadronisation in proton--proton (pp) collisions. In proton--lead (p--Pb) collisions, the measurements of charm baryons provide important information about cold nuclear matter effects and help to understand how the possible presence of collective effects could modify the production of heavy-flavour hadrons. In this contribution, the most recent results on open charm-hadron production in pp and p--Pb collisions measured by ALICE are discussed.

}
\maketitle
\section{Introduction}
\label{intro}
The production of heavy-flavour hadrons in high-energy hadronic collisions can provide important tests of the theory of quantum chromodynamics (QCD). The production cross sections of heavy-flavour hadrons can be calculated using the factorisation approach as a convolution of three factors \cite{COLLINS198637}: the parton distribution functions (PDFs) of the incoming nuclei, the hard-scattering cross section at partonic level, calculated as a perturbative series in powers of the strong coupling constant $\alpha_{\rm s}$, and the fragmentation functions of heavy quarks into corresponding heavy-flavour hadrons, which is an inherently non-perturbative process related to, or even driven by, the confining property of QCD. The heavy-flavour baryon-to-meson ratio is an ideal observable related to the hadronization mechanism since the contributions from parton distribution function and parton-parton scattering terms cancel in the ratio. The $\rm \Lambda_c^+/D^0$ ratio at the LHC is enhanced with respect to predictions based on $\rm e^+e^-$ and $\rm ep$ experiments, suggesting that the charm fragmentation functions are not universal among different collision systems. Several hadronization mechanisms, such as colour reconnection (CR) beyond the leading colour approximation \cite{Christiansen:2015yqa}, coalescence \cite{Minissale:2020bif, Song:2018tpv} and feed-down from a largely augmented set of higher mass charm-baryon states beyond the current listing of the particle data group (PDG) \cite{He:2019tik}, have been proposed to explain this enhancement. The newest measurements of the charm baryons $\rm \Lambda_c^+$, $\rm \Sigma_c^{0, ++}$, $\rm \Xi_c^{0, +}$ and $\rm \Omega_c^{0}$ performed with the ALICE experiment will be used to verify predictions from these hadronization mechanisms.

\section{Charm baryon-to-meson yield ratios in pp collisions}
\label{sec_2}
\begin{figure}
\centering
\includegraphics[width=0.3\textwidth]{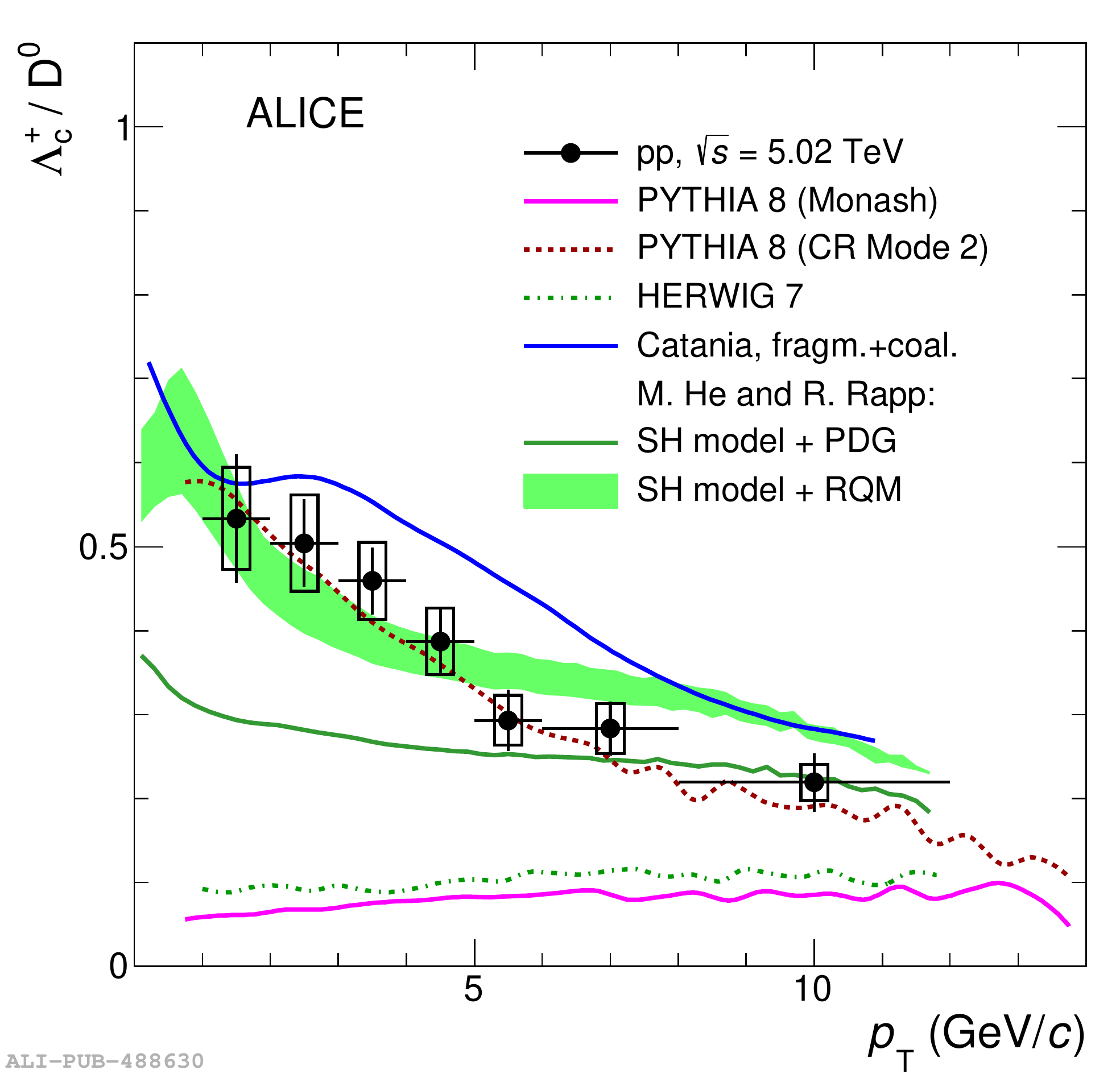}
\includegraphics[width=0.33\textwidth]{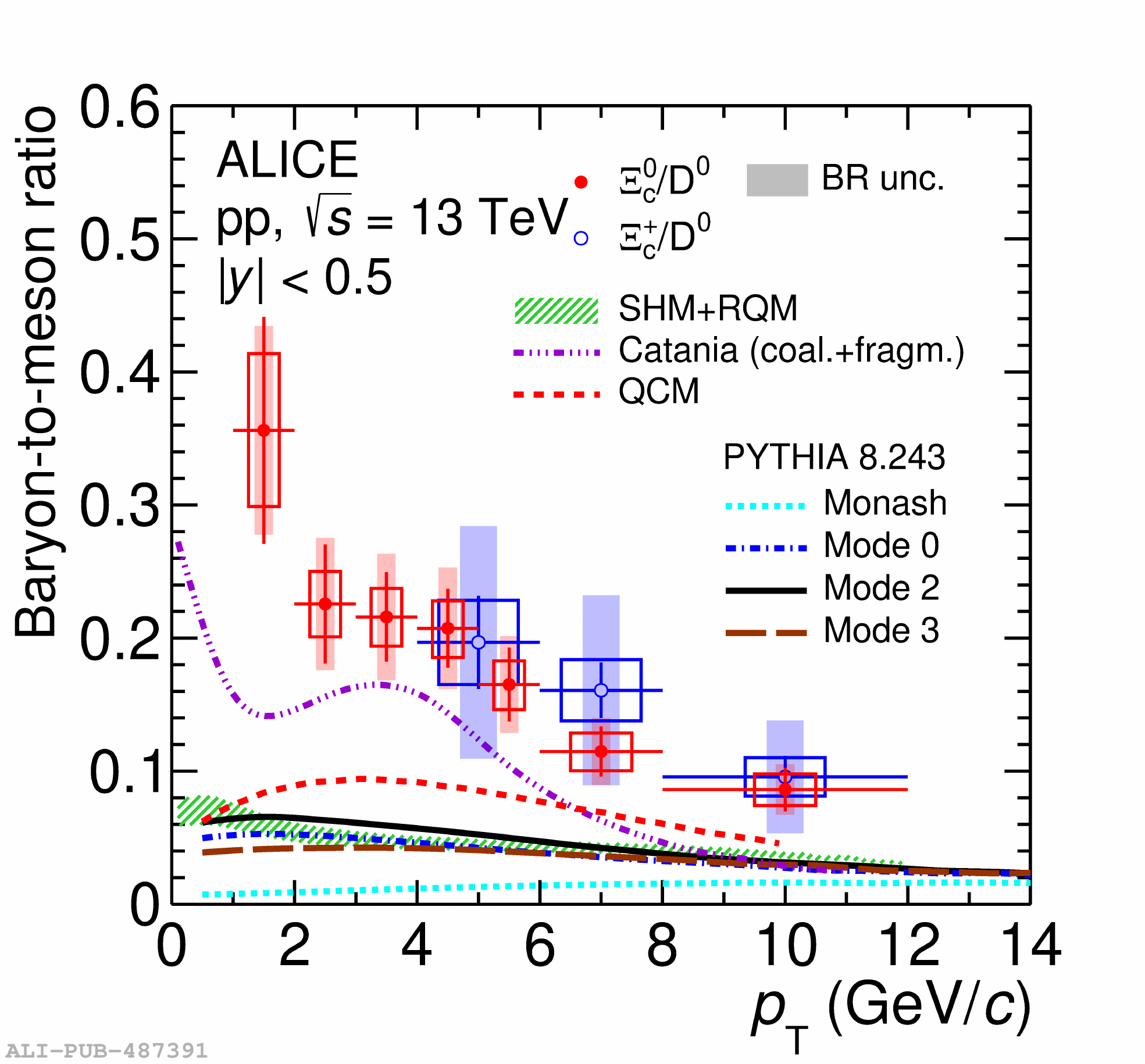}
\includegraphics[width=0.35\textwidth]{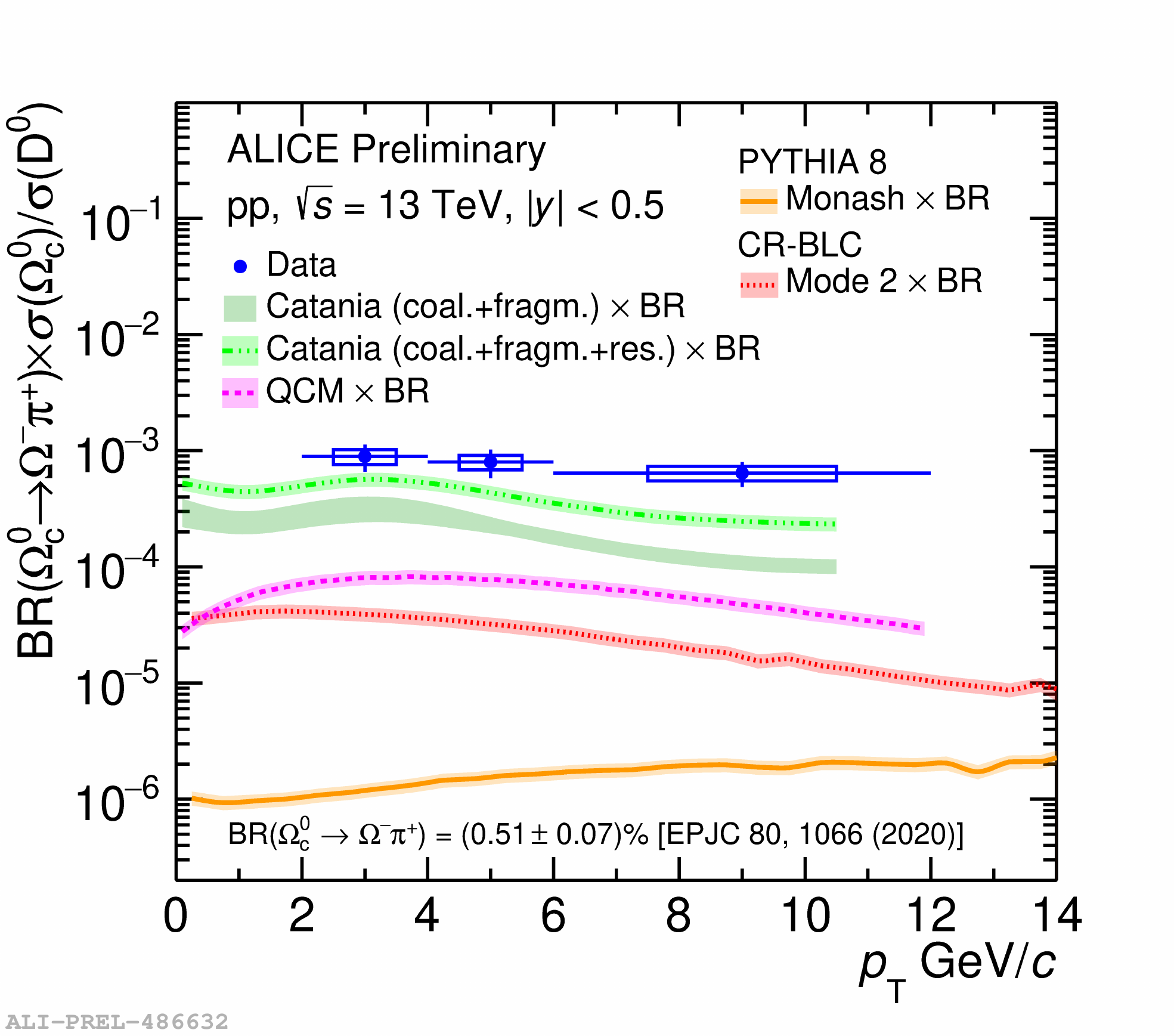}
\caption{Left: The $\rm \Lambda_c^+/D^0$ ratio as a function of $p_{\rm T}$ measured in pp collisions at $\sqrt{s}=5.02$ TeV \cite{ALICE:2020wla}. Middle: The $\rm \Xi_c^0/D^0$ and $\rm \Xi_c^+/D^0$ ratios measured in pp collisions at $\sqrt{s}=13$ TeV \cite{ALICE:2021bli}. Right: The $\rm BR\times\Omega_c^0/D^0$ measured in pp collisions at $\sqrt{s}=13$ TeV. All the charm baryon-to-meson yield ratios are compared to theoretical calculations \cite{Skands:2014pea, Christiansen:2015yqa, He:2019tik, Song:2018tpv, Minissale:2020bif}.}
\label{fig_1}       
\end{figure}

Thanks to the large data sample collected during the Run 2 period in pp collisions at $\sqrt{s}=5.02$ and 13 TeV, ALICE measured all the ground-state charm hadrons down to low transverse momentum ($p_{\rm T}$), including charm mesons ($\rm D^0$, $\rm D^+$, $\rm D_s^+$ \cite{ALICE:2021mgk}) and charm baryons ($\rm \Lambda_c^+$ \cite{ALICE:2020wla}, $\rm \Xi_c^{0,+}$ \cite{ALICE:2021bli, ALICE:2021psx}, and $\rm \Omega_c^0$). The $\rm \Lambda_c^+/D^0$ yield ratio is measured as a function of $p_{\rm T}$ in pp collisions at $\sqrt{s}=5.02$ TeV as shown in Fig.~\ref{fig_1} (left), the $\rm \Xi_c^{0}/D^0$ and $\rm \Xi_c^{+}/D^0$ yield ratios are measured in pp collisions at $\sqrt{s}=13$ TeV as shown in Fig.~\ref{fig_1} (middle). For $\rm \Omega_c^0$, the absolute decay branching ratio (BR) of $\rm \Omega_c^0\rightarrow\Omega^-\pi^+$ is not measured, hence the BR of $\rm \Omega_c^0\rightarrow\Omega^-\pi^+$ times cross section of $\rm \Omega_c^0$ over cross section of $\rm D^0$ in pp collisions at $\sqrt{s}=13$ TeV is reported in Fig.~\ref{fig_1} (right). In order to compare data with models, a theoretical calculation of BR($\rm \Omega_c^0\rightarrow\Omega^-\pi^+$) \cite{Hsiao:2020gtc} is used to multiply different models. The $\rm \Lambda_c^+/D^0$ and $\rm \Xi_c^0/D^0$ ratios show a downward trend with increasing $p_{\rm T}$. The Monte Carlo generator PYTHIA8 (Monash) \cite{Skands:2014pea} tuned on measurements in $\rm e^+e^-$ collisions largely underestimates all four charm baryon-to-meson yield ratios, providing evidence of different charm hadronisation mechanisms between $\rm e^+e^-$ and pp collisions. The $\rm \Lambda_c^+/D^0$ ratio is better described by a model with colour reconnection beyond the leading colour approximation \cite{Christiansen:2015yqa}, a statistical hadronisation model with an augmented set of charm baryon states predicted by the relativistic quark model (RQM) \cite{He:2019tik}, or a model relying on hadronisation via coalescence and fragmentation \cite{Minissale:2020bif}. However, all the models underestimate the $\rm \Xi_c^{0,+}/D^0$ and $\rm BR\times\Omega_c^0/D^0$ ratios, except the Catania model \cite{Minissale:2020bif} including charm quark hadronisation via both coalescence and fragmentation, which would indicate a partonic system similar to a quark--gluon plasma (QGP) in pp collisions.

\section{Charm production and fragmentation in pp collisions}
\label{sec_3}
\begin{figure}
\centering
\includegraphics[width=0.4\textwidth]{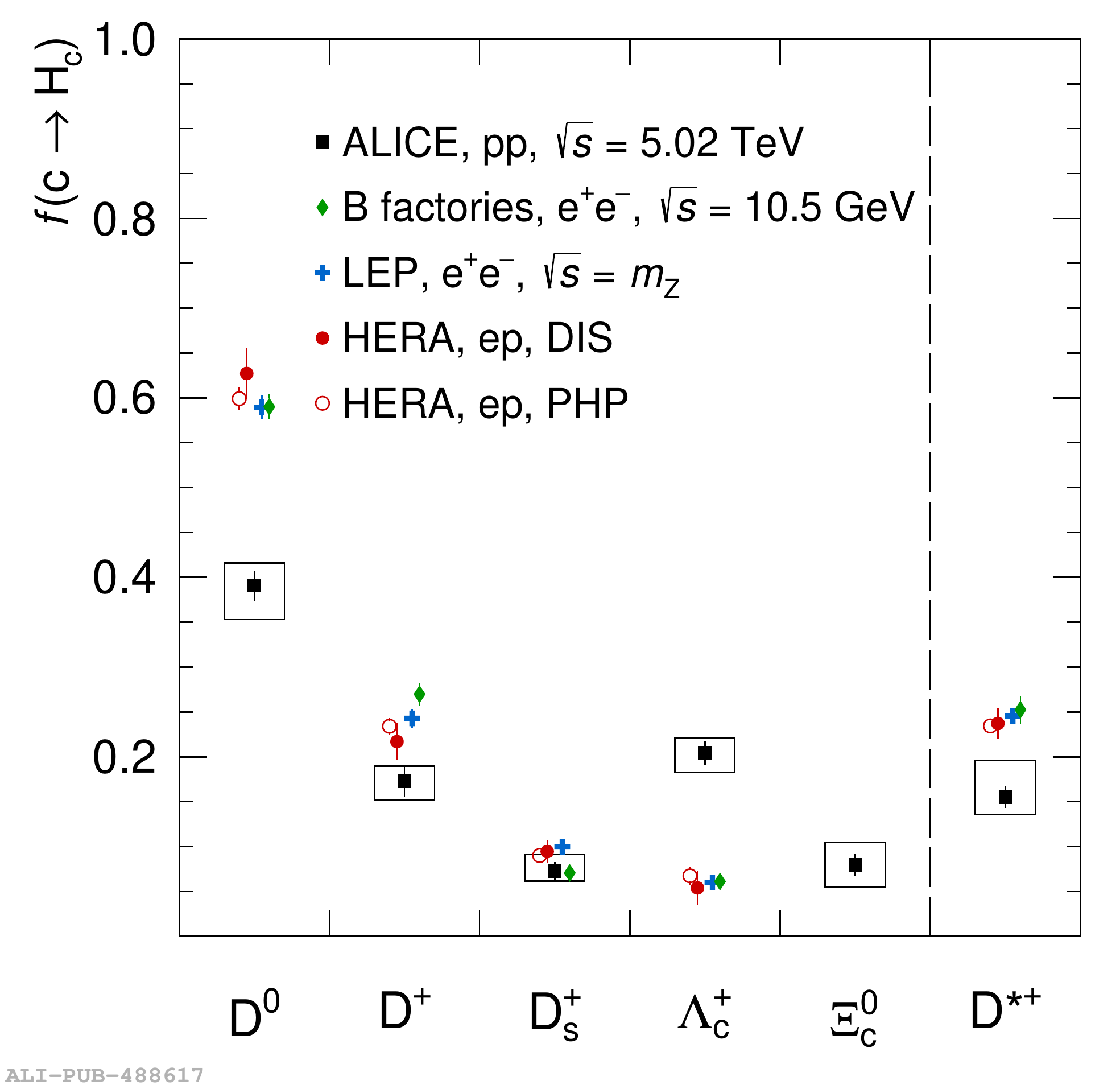}
\includegraphics[width=0.4\textwidth]{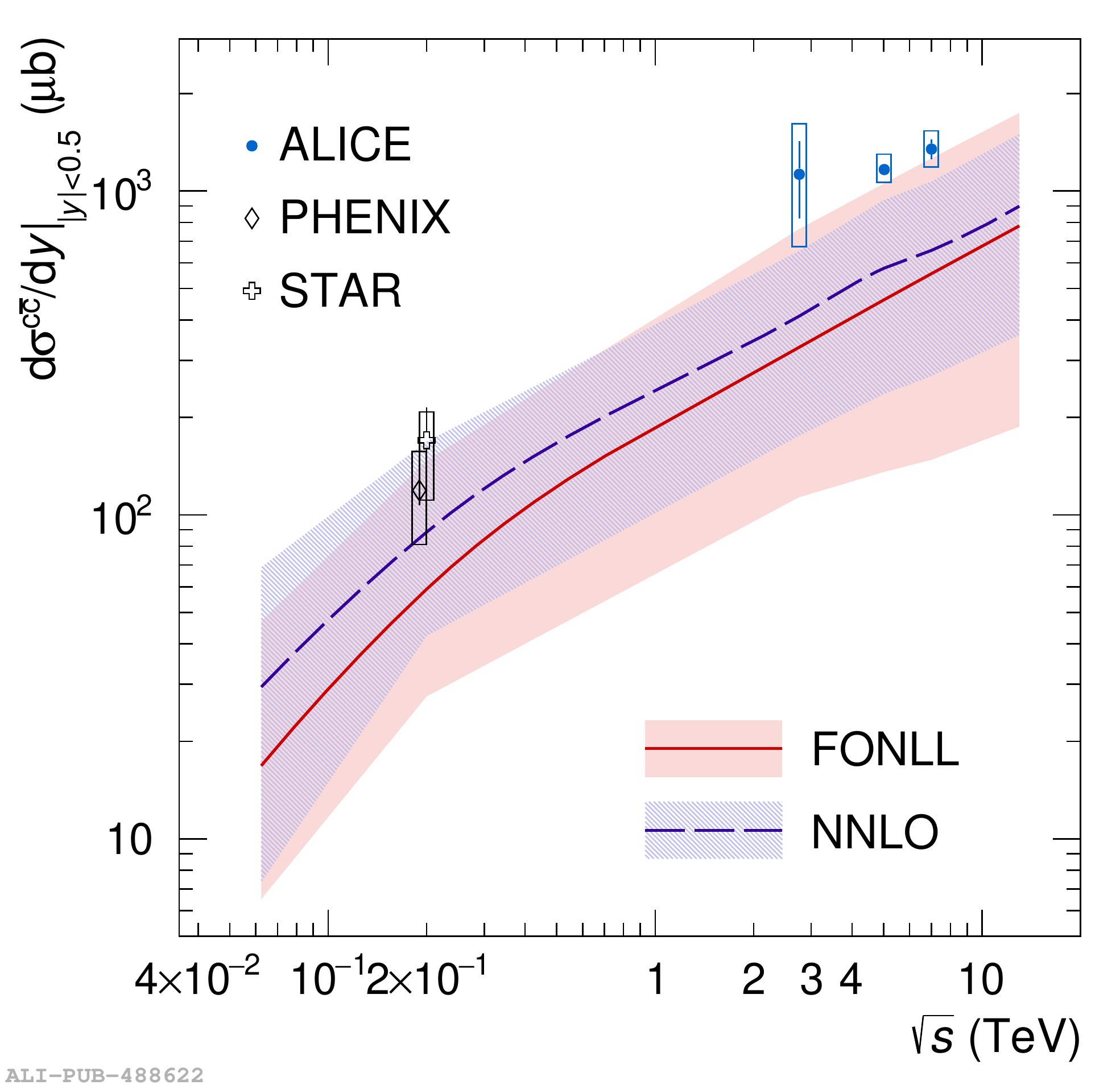}
\caption{Left: Charm-quark fragmentation fractions into charm hadrons measured in pp collisions at $\sqrt{s}=5.02$ TeV in comparison with experimental measurements performed in $\rm e^+e^-$ and ep collisions \cite{ALICE:2021dhb}. Right: Charm production cross section at midrapidity per unit of rapidity as a function of the collision energy at the LHC \cite{ALICE:2021dhb} and RHIC \cite{STAR:2012nbd, PHENIX:2010xji} compared to FONLL \cite{Cacciari:2012ny} and NNLO \cite{dEnterria:2016ids} calculations.}
\label{fig_2}       
\end{figure}

The charm fragmentation fraction $f(\rm c\rightarrow H_c)$ shown in Fig.~\ref{fig_2} (left) represents the probability of a charm quark hadronising into a given charm hadron. The fragmentation fraction for the $\rm \Xi_c^0$ baryon is the first measurement in any collision system. An increase of about a factor of 3.3 for the fragmentation fraction for the $\rm \Lambda_c^+$ baryon with respect to $\rm e^+e^-$ and ep collisions, and a corresponding decrease of about a factor of 1.2--1.4 for the $\rm D^0$ meson are observed, showing that the assumption of the charm fragmentation universality (collision-system independence) is broken. Charm quarks hadronise into baryons almost 40\% of the time, which is four times more often than what was measured at colliders with electron beams.

The $\rm c\bar{c}$ production cross section per unit of rapidity at midrapidity ($\rm d{\it \sigma}^{c\bar{c}}/d{\it y}|_{|{\it y}| < 0.5}$) is calculated by summing the $p_{\rm T}$-integrated cross sections of all measured ground-state charm hadrons ($\rm D^0$, $\rm D^+$, $\rm D_s^+$, $\rm \Lambda_c^+$, $\rm \Xi_c^0$ and their charge conjugates). The contribution of $\rm \Xi_c^0$ is multiplied by a factor of 2 in order to account for the contribution of $\rm \Xi_c^+$. Since the absence of a $\rm \Omega_c^0$ production measurement at hadron colliders, an asymmetric systematic uncertainty is assigned assuming a contribution equal to the one of $\rm \Xi_c^0$ considering the prediction of the Catania model \cite{Minissale:2020bif}. The resulting $\rm c\bar{c}$ production cross section per unit of rapidity at midrapidity is $\rm d{\it \sigma}^{c\bar{c}}/\rm d {\it y}|_{|{\it y}| < 0.5}^{\text{pp, 5.02 TeV}} = 1165\pm44\text{(stat)}^{+ 134}_{- 101}\text{(syst)}~{\rm \mu b}$. The updated fragmentation fractions obtained in pp collisions at $\sqrt{s}=5.02$ TeV allow the recomputation of the charm production cross sections per unit of rapidity at midrapidity in pp collisions at $\sqrt{s}=2.76$ and 7 TeV, which are about 40\% higher than the previously published results \cite{ALICE:2012inj, ALICE:2017olh}. The measured $\rm c\bar{c}$ production cross section per unit of rapidity at midrapidity together with measurements at RHIC \cite{STAR:2012nbd, PHENIX:2010xji} are located at the upper edge of FONLL \cite{Cacciari:2012ny} and NNLO \cite{dEnterria:2016ids} predictions.

\section{Charm hadronisation in p--Pb collisions}
\label{sec_4}
\begin{figure}
\centering
\includegraphics[width=0.37\textwidth]{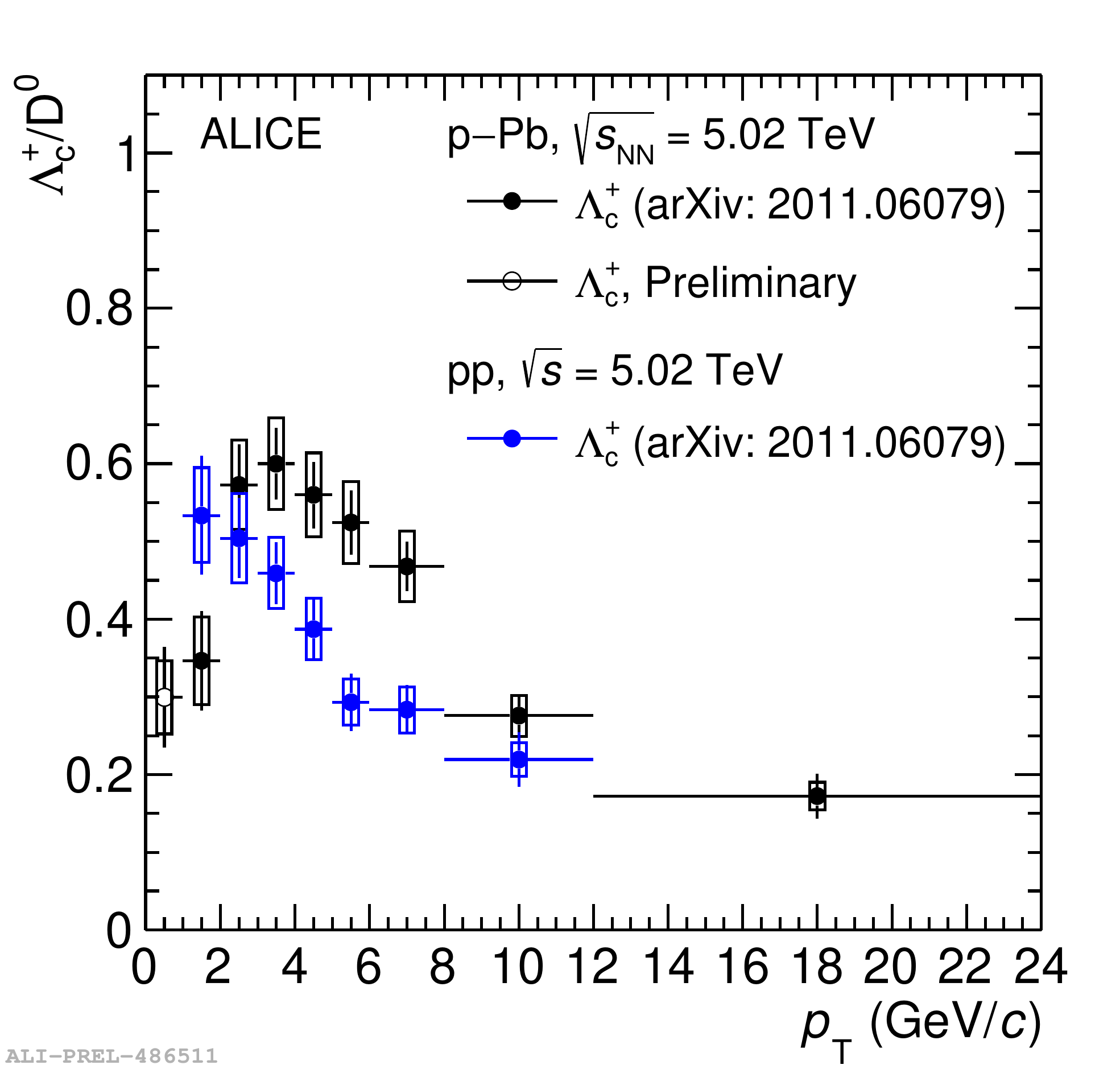}
\includegraphics[width=0.61\textwidth]{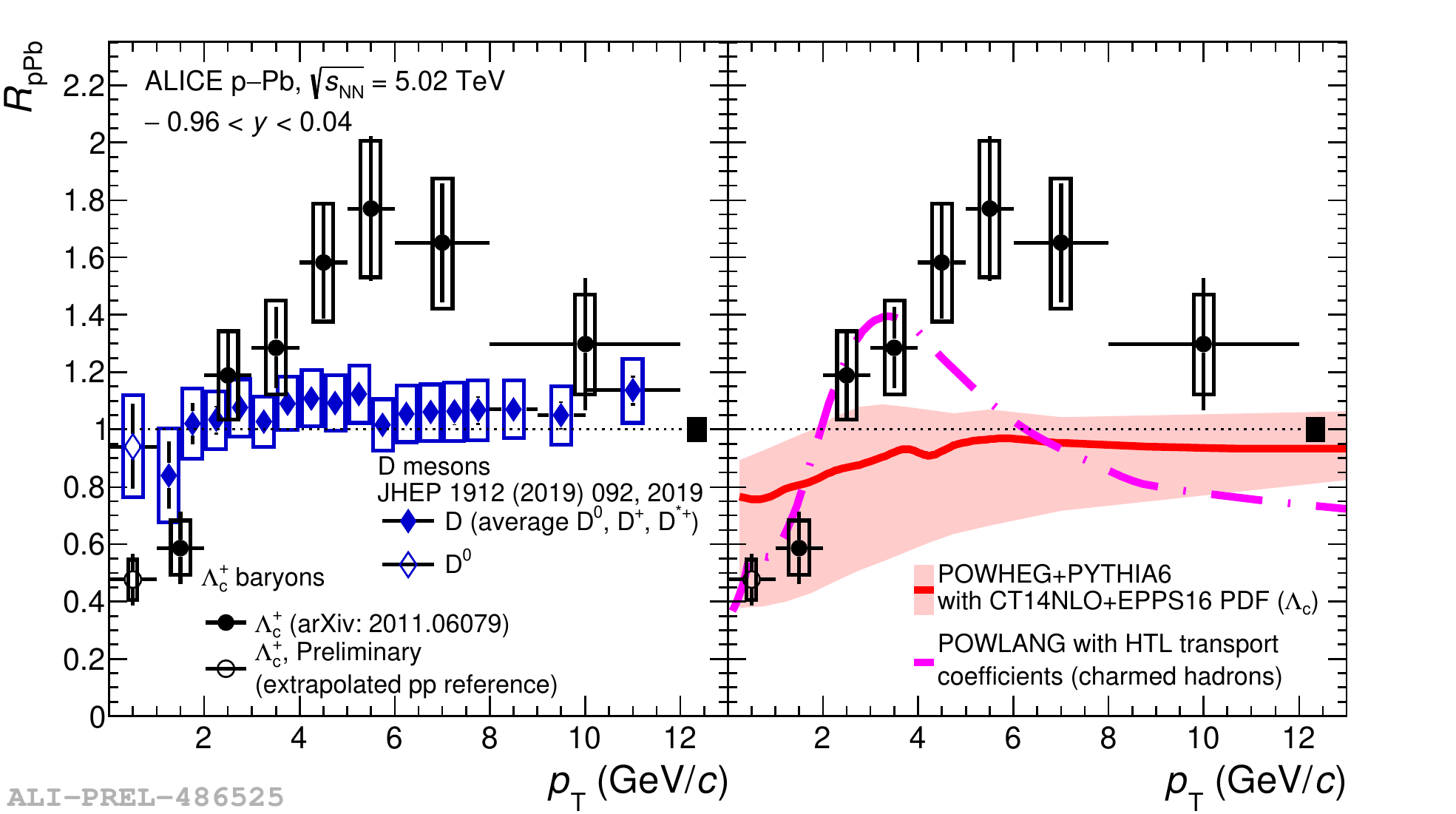}
\caption{Left: The $\rm \Lambda_c^+/D^0$ ratio as a function of $p_{\rm T}$ measured in pp collisions and p--Pb collisions at 5.02 TeV \cite{ALICE:2020wla}. Right: The nuclear modification factor $R_{\rm pPb}$ of the $\rm \Lambda_c^+$ baryon as a function of $p_{\rm T}$ in p--Pb collisions at $\sqrt{s_{\rm NN}}=5.02$ TeV, compared to $R_{\rm pPb}$ of D mesons, as well as to model expectations \cite{ALICE:2020wla}.}
\label{fig_3}       
\end{figure}

Charm hadronisation in p--Pb collisions at $\sqrt{s_{\rm NN}}=5.02$ TeV is also investigated. The left panel of Fig.~\ref{fig_3} shows the $\rm \Lambda_c^+/D^0$ ratio as a function of $p_{\rm T}$ in pp and p--Pb collisions at 5.02 TeV. This ratio is measured down to $p_{\rm T}=0$ in p--Pb collisions for the first time at the LHC. The measurements of $\rm \Lambda_c^+/D^0$ in pp and p--Pb collisions are qualitatively consistent with each other, although a larger ratio in $3<p_{\rm T}<8$ GeV/$c$ and a lower ratio in $1<p_{\rm T}<2$ GeV/$c$ are measured in p--Pb collisions with respect to pp collisions. A clear decreasing trend with increasing $p_{\rm T}$ is obtained in both pp and p--Pb collisions for $p_{\rm T}>2$ GeV/$c$. The nuclear modification factor $R_{\rm pPb}$ of the $\rm \Lambda_c^+$ baryon measured as a function of $p_{\rm T}$ together with $R_{\rm pPb}$ of non-strange D mesons are shown in the right panel of Fig.~\ref{fig_3}. There is a significant suppression for $p_{\rm T}<2$ GeV/$c$ and enhancement for $4<p_{\rm T}<8$ GeV/$c$ in p--Pb collisions with respect to pp collisions, which is similar as the $p_{\rm T}$ distribution of $\rm \Lambda_c^+/D^0$ ratio, suggesting the presence of possible radial flow effects or a further modification of the charm hadronisation mechanism in p--Pb collisions. The measurement of $R_{\rm pPb}$ for the $\rm \Lambda_c^+$ baryon is compared to model calculations.  The POWHEG \cite{Frixione:2007nw} + PYTHIA6 simulations use the POWHEG event generator with PYTHIA6 parton shower and EPPS16 \cite{Eskola:2016oht} parameterisation of the nuclear modification of the PDFs. The POWLANG model \cite{Beraudo:2015wsd} assumes that a hot deconfined medium is formed in p--Pb collisions. The two models capture some features of the data, but neither of them can quantitatively reproduce the data in the measured $p_{\rm T}$ interval.

\section{Acknowledgements}
This work was supported by the National Natural Science Foundation of China (NSFC) (No. 12105109).

%
%
%
\let\oldthebibliography\thebibliography
\let\endoldthebibliography\endthebibliography
\renewenvironment{thebibliography}[1]{
  \begin{oldthebibliography}{#1}
    \setlength{\itemsep}{0.03em}
    \setlength{\parskip}{0em}
}
{
  \end{oldthebibliography}
}

\bibliographystyle{utphys}
\bibliography{references}

\end{document}